\documentclass[letterpaper,aps,prl,twocolumn,superscriptaddress,amsmath, amssymb,showpacs]{revtex4}

\usepackage{color}
\usepackage{eso-pic}

\newif\ifPDF
 \ifx\pdfoutput\undefined
 \pdffalse                  
\else
 \pdfoutput=0               
 \PDFfalse
\fi

\ifPDF
    \usepackage[pdftex]{graphicx}
    \usepackage{epstopdf}
 \usepackage{url,boxedminipage}
 \usepackage{graphicx,thumbpdf}
 \definecolor{rltred}{rgb}{0.75,0,0}
 \definecolor{rltgreen}{rgb}{0,0.5,0}
 \definecolor{rltblue}{rgb}{0,0,0.75}

 \usepackage[colorlinks%
  ,bookmarks%
  ,urlcolor=rltblue%
  ,filecolor=rltgreen%
  ,linkcolor=rltred%
  ,pdftitle={pdftitLe}%
  ,pdfsubject={pdfsubject}%
  ,pdfkeywords={pdfkeywords}%
  ,pdfauthor={Joe User <joeuser@fnal.gov>}%
  ,pdfpagemode={UseOutlines}%
  ,bookmarksopen=true%
  ,bookmarksnumbered=true%
  ,pdfstartview={Fit}%
  ]{hyperref}
 \DeclareGraphicsExtensions{.pdf,.jpg,.jpeg}
\else
 \usepackage{graphicx}
 \DeclareGraphicsExtensions{.eps,.ps,.eps.gz,.ps.gz}
 \newcommand{\href}[2]{#2}                   
 \usepackage[dvips,pagebackref]{hyperref}    
\fi 

\usepackage{dcolumn}
\usepackage{bm}
\usepackage{setspace}
\usepackage{multirow}

\newcommand{\ppbar}{\ensuremath {p\overline{p}}}

\newcommand{\Zee}{\ensuremath{Z \rightarrow e^+e^-}}
\newcommand{\Zll}{\ensuremath{Z \rightarrow l^+l^-}}

\newcommand{\Zllg}{\ensuremath{Z \gamma \rightarrow l^+l^- \gamma}}

\newcommand{\Znn}{\ensuremath{Z \rightarrow \nu \bar{\nu}}}

\newcommand{\Zmm}{\ensuremath{Z \rightarrow \mu^+ \mu^-}}

\newcommand{\Zg}{\ensuremath{Z\gamma}}

\newcommand{\GeV}{\ensuremath{\mathrm{Ge\kern -0.1em V}}}
\newcommand{\GeVc}{\ensuremath{\GeV\!/c}}
\newcommand{\GeVcc}{\ensuremath{\GeV\!/c^2}}

\newcommand{\invfb}{\ensuremath{\mathrm{fb^{-1}}}}

\newcommand{\Pt}{\ensuremath {p_{{T}}}}
\def\sla#1{\rlap{\kern .15em /}#1}
\newcommand{\met}{\ensuremath{\sla{E}_T}}
\newcommand{\Et}{\ensuremath{E_{T}}}
\newcommand{\Etg}{\ensuremath{E_{T}^{\gamma}}}

\begin{document}

%

\preprint{CDF/PUB/ELECTROWEAK/CDFR/10375}
\preprint{FERMILAB-PUB-11-065-E}


\title{Limits on Anomalous Trilinear Gauge Couplings in {\boldmath $Z\gamma$} Events from \ppbar\ Collisions at $\sqrt{s} =$ 1.96 TeV}

\affiliation{Institute of Physics, Academia Sinica, Taipei, Taiwan 11529, Republic of China} 
\affiliation{Argonne National Laboratory, Argonne, Illinois 60439, USA} 
\affiliation{University of Athens, 157 71 Athens, Greece} 
\affiliation{Institut de Fisica d'Altes Energies, ICREA, Universitat Autonoma de Barcelona, E-08193, Bellaterra (Barcelona), Spain} 
\affiliation{Baylor University, Waco, Texas 76798, USA} 
\affiliation{Istituto Nazionale di Fisica Nucleare Bologna, $^{aa}$University of Bologna, I-40127 Bologna, Italy} 
\affiliation{University of California, Davis, Davis, California 95616, USA} 
\affiliation{University of California, Los Angeles, Los Angeles, California 90024, USA} 
\affiliation{Instituto de Fisica de Cantabria, CSIC-University of Cantabria, 39005 Santander, Spain} 
\affiliation{Carnegie Mellon University, Pittsburgh, Pennsylvania 15213, USA} 
\affiliation{Enrico Fermi Institute, University of Chicago, Chicago, Illinois 60637, USA}
\affiliation{Comenius University, 842 48 Bratislava, Slovakia; Institute of Experimental Physics, 040 01 Kosice, Slovakia} 
\affiliation{Joint Institute for Nuclear Research, RU-141980 Dubna, Russia} 
\affiliation{Duke University, Durham, North Carolina 27708, USA} 
\affiliation{Fermi National Accelerator Laboratory, Batavia, Illinois 60510, USA} 
\affiliation{University of Florida, Gainesville, Florida 32611, USA} 
\affiliation{Laboratori Nazionali di Frascati, Istituto Nazionale di Fisica Nucleare, I-00044 Frascati, Italy} 
\affiliation{University of Geneva, CH-1211 Geneva 4, Switzerland} 
\affiliation{Glasgow University, Glasgow G12 8QQ, United Kingdom} 
\affiliation{Harvard University, Cambridge, Massachusetts 02138, USA} 
\affiliation{Division of High Energy Physics, Department of Physics, University of Helsinki and Helsinki Institute of Physics, FIN-00014, Helsinki, Finland} 
\affiliation{University of Illinois, Urbana, Illinois 61801, USA} 
\affiliation{The Johns Hopkins University, Baltimore, Maryland 21218, USA} 
\affiliation{Institut f\"{u}r Experimentelle Kernphysik, Karlsruhe Institute of Technology, D-76131 Karlsruhe, Germany} 
\affiliation{Center for High Energy Physics: Kyungpook National University, Daegu 702-701, Korea; Seoul National University, Seoul 151-742, Korea; Sungkyunkwan University, Suwon 440-746, Korea; Korea Institute of Science and Technology Information, Daejeon 305-806, Korea; Chonnam National University, Gwangju 500-757, Korea; Chonbuk National University, Jeonju 561-756, Korea} 
\affiliation{Ernest Orlando Lawrence Berkeley National Laboratory, Berkeley, California 94720, USA} 
\affiliation{University of Liverpool, Liverpool L69 7ZE, United Kingdom} 
\affiliation{University College London, London WC1E 6BT, United Kingdom} 
\affiliation{Centro de Investigaciones Energeticas Medioambientales y Tecnologicas, E-28040 Madrid, Spain} 
\affiliation{Massachusetts Institute of Technology, Cambridge, Massachusetts 02139, USA} 
\affiliation{Institute of Particle Physics: McGill University, Montr\'{e}al, Qu\'{e}bec, Canada H3A~2T8; Simon Fraser University, Burnaby, British Columbia, Canada V5A~1S6; University of Toronto, Toronto, Ontario, Canada M5S~1A7; and TRIUMF, Vancouver, British Columbia, Canada V6T~2A3} 
\affiliation{University of Michigan, Ann Arbor, Michigan 48109, USA} 
\affiliation{Michigan State University, East Lansing, Michigan 48824, USA}
\affiliation{Institution for Theoretical and Experimental Physics, ITEP, Moscow 117259, Russia}
\affiliation{University of New Mexico, Albuquerque, New Mexico 87131, USA} 
\affiliation{Northwestern University, Evanston, Illinois 60208, USA} 
\affiliation{The Ohio State University, Columbus, Ohio 43210, USA} 
\affiliation{Okayama University, Okayama 700-8530, Japan} 
\affiliation{Osaka City University, Osaka 588, Japan} 
\affiliation{University of Oxford, Oxford OX1 3RH, United Kingdom} 
\affiliation{Istituto Nazionale di Fisica Nucleare, Sezione di Padova-Trento, $^{bb}$University of Padova, I-35131 Padova, Italy} 
\affiliation{LPNHE, Universite Pierre et Marie Curie/IN2P3-CNRS, UMR7585, Paris, F-75252 France} 
\affiliation{University of Pennsylvania, Philadelphia, Pennsylvania 19104, USA}
\affiliation{Istituto Nazionale di Fisica Nucleare Pisa, $^{cc}$University of Pisa, $^{dd}$University of Siena and $^{ee}$Scuola Normale Superiore, I-56127 Pisa, Italy} 
\affiliation{University of Pittsburgh, Pittsburgh, Pennsylvania 15260, USA} 
\affiliation{Purdue University, West Lafayette, Indiana 47907, USA} 
\affiliation{University of Rochester, Rochester, New York 14627, USA} 
\affiliation{The Rockefeller University, New York, New York 10065, USA} 
\affiliation{Istituto Nazionale di Fisica Nucleare, Sezione di Roma 1, $^{ff}$Sapienza Universit\`{a} di Roma, I-00185 Roma, Italy} 

\affiliation{Rutgers University, Piscataway, New Jersey 08855, USA} 
\affiliation{Texas A\&M University, College Station, Texas 77843, USA} 
\affiliation{Istituto Nazionale di Fisica Nucleare Trieste/Udine, I-34100 Trieste, $^{gg}$University of Trieste/Udine, I-33100 Udine, Italy} 
\affiliation{University of Tsukuba, Tsukuba, Ibaraki 305, Japan} 
\affiliation{Tufts University, Medford, Massachusetts 02155, USA} 
\affiliation{University of Virginia, Charlottesville, VA  22906, USA}
\affiliation{Waseda University, Tokyo 169, Japan} 
\affiliation{Wayne State University, Detroit, Michigan 48201, USA} 
\affiliation{University of Wisconsin, Madison, Wisconsin 53706, USA} 
\affiliation{Yale University, New Haven, Connecticut 06520, USA} 
\author{T.~Aaltonen}
\affiliation{Division of High Energy Physics, Department of Physics, University of Helsinki and Helsinki Institute of Physics, FIN-00014, Helsinki, Finland}
\author{B.~\'{A}lvarez~Gonz\'{a}lez$^w$}
\affiliation{Instituto de Fisica de Cantabria, CSIC-University of Cantabria, 39005 Santander, Spain}
\author{S.~Amerio}
\affiliation{Istituto Nazionale di Fisica Nucleare, Sezione di Padova-Trento, $^{bb}$University of Padova, I-35131 Padova, Italy} 

\author{D.~Amidei}
\affiliation{University of Michigan, Ann Arbor, Michigan 48109, USA}
\author{A.~Anastassov}
\affiliation{Northwestern University, Evanston, Illinois 60208, USA}
\author{A.~Annovi}
\affiliation{Laboratori Nazionali di Frascati, Istituto Nazionale di Fisica Nucleare, I-00044 Frascati, Italy}
\author{J.~Antos}
\affiliation{Comenius University, 842 48 Bratislava, Slovakia; Institute of Experimental Physics, 040 01 Kosice, Slovakia}
\author{G.~Apollinari}
\affiliation{Fermi National Accelerator Laboratory, Batavia, Illinois 60510, USA}
\author{J.A.~Appel}
\affiliation{Fermi National Accelerator Laboratory, Batavia, Illinois 60510, USA}
\author{A.~Apresyan}
\affiliation{Purdue University, West Lafayette, Indiana 47907, USA}
% ...
\collaboration{CDF Collaboration\footnote{With visitors from $^a$University of Massachusetts Amherst, Amherst, Massachusetts 01003,
$^b$Istituto Nazionale di Fisica Nucleare, Sezione di Cagliari, 09042 Monserrato (Cagliari), Italy,
$^c$University of California Irvine, Irvine, CA  92697, 
$^d$University of California Santa Barbara, Santa Barbara, CA 93106
$^e$University of California Santa Cruz, Santa Cruz, CA  95064,
$^f$CERN,CH-1211 Geneva, Switzerland,
$^g$Cornell University, Ithaca, NY  14853, 
$^h$University of Cyprus, Nicosia CY-1678, Cyprus, 
$^i$University College Dublin, Dublin 4, Ireland,
$^j$University of Fukui, Fukui City, Fukui Prefecture, Japan 910-0017,
$^k$Universidad Iberoamericana, Mexico D.F., Mexico,
$^l$Iowa State University, Ames, IA  50011,
$^m$University of Iowa, Iowa City, IA  52242,
$^n$Kinki University, Higashi-Osaka City, Japan 577-8502,
$^o$Kansas State University, Manhattan, KS 66506,
$^p$University of Manchester, Manchester M13 9PL, England,
$^q$Queen Mary, University of London, London, E1 4NS, England,
$^r$University of Melbourne, Victoria 3010, Australia,
$^s$Muons, Inc., Batavia, IL 60510,
$^t$Nagasaki Institute of Applied Science, Nagasaki, Japan, 
$^u$National Research Nuclear University, Moscow, Russia,
$^v$University of Notre Dame, Notre Dame, IN 46556,
$^w$Universidad de Oviedo, E-33007 Oviedo, Spain, 
$^x$Texas Tech University, Lubbock, TX  79609, 
$^y$Universidad Tecnica Federico Santa Maria, 110v Valparaiso, Chile,
$^z$Yarmouk University, Irbid 211-63, Jordan,
$^{hh}$On leave from J.~Stefan Institute, Ljubljana, Slovenia
}}
\noaffiliation

\date{\today}

\begin{abstract}

Using \Zg\ candidate events collected by the CDF detector at the Tevatron Collider, we search for potential anomalous (non-standard-model) couplings between the $Z$ boson and the photon. \Zg\ couplings vanish at tree level and are heavily suppressed at higher orders; hence any evidence of couplings indicates new physics. Measurements are performed using data corresponding to an integrated luminosity of 4.9 \invfb\ in the $Z \rightarrow \nu\bar{\nu}$ decay channel and 5.1 \invfb\ in the \Zll\ ($l=\mu, e$) decay channels. The combination of these measurements provides the most stringent limits to date on \Zg\ trilinear gauge couplings. Using an energy scale of $\Lambda = 1.5$\ TeV to allow for a direct comparison with previous measurements, we find limits on the CP-conserving parameters that describe \Zg\ couplings to be $|h_3^{\gamma,Z}| < 0.022 $ and $|h_4^{\gamma,Z}| < 0.0009$. These results are consistent with standard model predictions.

\end{abstract}


\maketitle


Studies of trilinear couplings between the gauge bosons ($W, Z, \gamma$) test the standard model (SM) description of gauge sector interactions and provide sensitivity to physics beyond the SM through examination of production rates and kinematics~\cite{CDFZllAGC,CDFWWAGC,D0ZllAGC,D0ZnunuAGC,D0WZAGC, LEPEWWG}. In the case of neutral couplings, $ZZ\gamma$ and $Z\gamma\gamma$ vertex interactions vanish at tree level and, while allowed via internal particle loops, are highly suppressed in the SM. However, these trilinear gauge couplings can be non-negligible if loop contributions occur via non-SM particles. Models such as those incorporating compositeness or supersymmetry can alter the predicted cross section and production kinematics of \Zg\ events~\cite{newZgPhys,tgc,prd41_1990_1476,prd47_1993_4889}. 

In the SM, given the suppression of $ZZ\gamma$ and $Z\gamma\gamma$ couplings, the production of \Zg\ events is dominated by production of a $Z$ boson along with the radiation of a photon off either an incoming parton or a $Z$ decay product. These production mechanisms are interesting in their own right, serving as an important background 
to searches for new physics (e.g.~in gauge-mediated supersymmetry breaking models~\cite{gmsb}) and Higgs boson searches.
In this Letter, the production properties of \Zg\ events are compared 
to SM predictions, and limits are set on anomalous trilinear gauge couplings.  

The measurements of $Z \gamma$ couplings are performed with
\ppbar\ collision data at $\sqrt{s}$ = 1.96 TeV from the Tevatron Collider using the Collider Detector at Fermilab (CDF). 
We seek two types of \Zg\ events: those where the $Z$ decays to charged leptons (by identifying
lepton candidate pairs and a prompt photon~\cite{photonID} with large transverse energy \Et~\cite{coords}), and those where the $Z$ decays to
neutrinos (by identifying an event with only a solitary, prompt, high-\Et\ photon). In the former case, data corresponding to an integrated luminosity of 5.1 \invfb\ are used; in the latter, 4.9 \invfb . These measurements use over twice as much data as the previous published CDF result~\cite{CDFZllAGC} and incorporate looser muon selection requirements. As no significant disagreement is found between the SM prediction and the data, we set limits that are not only far more restrictive than those measured in~\cite{CDFZllAGC}, but are approximately half the magnitude of the previous best published limits~\cite{D0ZnunuAGC}. 

In beyond-the-SM scenarios with enhanced \Zg\
couplings, not only does the \Zg\ production cross section increase, but the photon \Et\ spectrum is modified due to an enhancement in the production of high-\Et\ photons~\cite{prd47_1993_4889}. We take advantage of this enhancement by
comparing the photon \Et\ distribution in data to both SM and beyond-the-SM predictions. Binned maximum likelihood measurements of the coupling parameters that describe \Zg\ interactions in the Lagrangian are performed. We calculate separate likelihoods for the \Zll\ and \Znn\ samples and combine the likelihoods to produce the final result. 

The CDF detector is covered in detail elsewhere~\cite{CDF,CDF_run2}. The transverse momenta (\Pt) of charged particles
are measured by an eight-layer silicon strip detector~\cite{SVX} 
and a 96-layer drift chamber (COT)~\cite{COT} inside a 1.4\,T magnetic field.
The COT provides tracking coverage with high efficiency for the pseudorapidity range $|\eta|<1$~\cite{coords}. 
Electromagnetic and hadronic calorimeters
surround the tracking system. They are segmented in a projective tower geometry and measure
the energies of charged and neutral particles in the central ($|\eta|<1.1$) and forward ($1.1<|\eta|<3.6$) regions. 
Each calorimeter has an electromagnetic shower profile detector 
positioned at the shower maximum~\cite{showerMax}.  
The calorimeters are surrounded by drift chambers that detect muons.

The measurements of anomalous trilinear gauge coupling parameters in the \Zll\ and $Z \rightarrow \nu\bar{\nu}$ decay channels differ both in event selection and background estimation. For the \Zllg\ decay channel we identify events containing \Zmm\ and \Zee\ candidates along with prompt photon candidates with $\Etg > 50$ GeV. According to experiments performed on simulated events, this choice of \Etg\ requirement maximizes the ability of the analysis to exclude anomalous couplings assuming SM physics, although a serious loss in sensitivity only occurs if the \Etg\ requirement is placed at 100 GeV or higher. The previous CDF analysis used a much less restrictive requirement of $\Etg > 7$ GeV, as the \Zg\ cross section was being measured in addition to trilinear gauge coupling parameters~\cite{CDFZllAGC}; additionally, placing the cut at 50 GeV allows for a control region to be based off of lower-$E_T$ photons. Event selection starts with inclusive muon (electron) triggers that require muon $\Pt > 18$ \GeVc\ (electron $\Et > 18 $ GeV). For electrons, a track must be reconstructed in the COT or in the silicon detector; additionally, the energy deposited by the candidate in the calorimeter must be isolated. For muons, a track must be reconstructed in the COT; additionally, no more than a few GeV of energy may be deposited in the calorimeters so that the candidate is compatible with a minimum ionizing particle. The two lepton candidates must correspond to the same flavor, with a requirement of $\Pt > 20$ \GeVc\ ($\Et > 20$ GeV) on one muon (electron) candidate and $\Pt > 10$ \GeVc\ ($\Et > 10$ GeV) on the other; furthermore, if the charges of both leptons are well-measured, the signs of these charges must be opposite. Studies of the invariant mass distributions of the two lepton candidates indicate that we retain a very high purity of $Z$ bosons (over $99\%$) despite the loose selection requirements. 

Once we have selected events with \Zll\ candidates, we look for isolated photons that pass standard CDF requirements~\cite{photonIsolation} in the central region $(|\eta| < 1.1)$ with $\Etg > 50$ GeV and are well separated from the $Z$ decay leptons ($\Delta R_{\ell\gamma} > 0.7$, with $\Delta R = \sqrt{ (\phi_l - \phi_{\gamma})^{2} + (\eta_l - \eta_{\gamma})^{2}}$). Additionally, we require that the two lepton candidates and the photon candidate form a three-body invariant mass greater than 100 \GeVcc\ in order to discriminate against events where the photon is radiated from one of the leptons from the $Z$ boson decay. The estimated contribution of SM $Z\gamma$ events is derived from Monte Carlo (MC) simulations that use the Baur-Berger package at the generator level~\cite{prd47_1993_4889} and \textsc{pythia}~\cite{pythia} for particle showering. This method yields a prediction of 87.2 $\pm$ 7.8 \Zg\ events that pass our selection requirements, where the uncertainty is dominated by the uncertainty on the luminosity and the predicted cross section. The non-\Zg\ events that pass these selection requirements result from hadronic jets being reconstructed as prompt photons and leptons (more commonly electrons). This background is estimated by calculating separate probabilities for a jet to mimic a photon or lepton as a function of jet \Et, and applying them to jets in events to which all our requirements have been applied except those pertaining to the mimicked particle. For photons and electrons, these probabilities are calculated by taking the ratio of the number of individual photon or electron candidates to the number of jets in a sample of data events where only the presence of at least one jet is required. The number of photon and electron candidates is corrected for the expected contribution of true photons or electrons in this sample. We estimate the probability for a false muon candidate from the number of dimuon $Z$ decay candidates in which both muon candidates have the same charge. Overall, the non-\Zg\ background contribution is very low: of the 91 events that pass our requirements, less than one event involving a mimicked photon or lepton is expected. 

	In order to identify $Z\gamma$ candidate events in the $Z \rightarrow \nu\bar{\nu}$ decay channel, we require solitary high-\Et\ photons and a transverse energy imbalance~\cite{MET_defn} in the detector. These events must pass a trigger requirement of an electromagnetic cluster with $\Et > 25$ GeV and $|\eta| < 1.1$ as well as missing transverse energy in excess of 25 GeV. For our signal region we require $\Etg > 100$ GeV, a threshold optimized in the same manner as the \Zll\ case. To account for the neutrinos we require a transverse energy imbalance of at least 50 GeV. In order to discriminate against $W$ boson contamination in our sample, we reject events containing any tracks with $\Pt > 10$ GeV, any electron candidates with $\Et > 15$ GeV, or any muon candidates with $\Pt > 10$ \GeVc. Additionally, we reject events that have any jets with $\Et > 15$ GeV in order to reduce the mismeasurement of missing transverse energy. The primary SM source for photons passing these requirements is $Z\gamma$ events in which the $Z$ has decayed to a pair of neutrinos, as shown in Table~\ref{tab:Znunu_bkgnds}. The method of estimating the expected number of $Z\gamma$ events is the same as that used for the \Zll\ candidate sample.   

The primary source of non-\Zg\ events in the final $Z\gamma \rightarrow \nu\bar{\nu}\gamma$ candidate sample is cosmic ray interactions. High-\Et\ photons from cosmic rays leave large transverse energy imbalances in our detector, mimicking the presence of neutrinos. Therefore, additional event requirements are applied to reduce the contributions from cosmic ray events. First, we require that  the energy deposited in the electromagnetic calorimeter appear within a timing window centered on the \ppbar\ interaction. Second, we use a relevance vector machine (RVM) multivariate discriminator~\cite{RVM} to distinguish whether a photon came from a collision or a non-collision source; the three inputs used for the RVM discriminator are the $\phi$-angle between the photon candidate and the closest muon candidate (if any), the ratio of energies from the photon candidate in the electromagnetic and hadronic calorimeters, and the ratio of energies from the electromagnetic shower profile detector and the electromagnetic calorimeter. We use photons outside the timing window to train the RVM for non-collision sources, and photons recoiling against jets to train for collision sources. The RVM discriminator reduces the contribution from cosmic ray events by an additional 90\%. Finally, we require the event to have a reconstructed vertex of at least three tracks from a \ppbar\ interaction. After applying these selection requirements, we have 85 candidate events in our sample. Despite the anti-cosmic ray requirements, cosmic ray events remain the second largest contributor to our sample, after \Zg\ events. 

We model two other major categories of non-\Zg\ events: one in which a charged lepton from $W \rightarrow e\nu$, $W \rightarrow \mu\nu$, or $W \rightarrow \tau\nu$ decay is reconstructed as a photon, and the other in which a true photon is produced but another object (e.g.~a lepton) is lost or only partially reconstructed, creating a large transverse energy imbalance. For the former case, the rate at which electrons are reconstructed as photons in the detector has been calculated using events with an electron and photon pair candidate that has an invariant mass near the mass of the $Z$, i.e., events in which the photon candidates are almost entirely electrons in actuality. The rate at which $\mu$'s and $\tau$'s are reconstructed as photons is taken from MC. For the latter case, which encompasses $W\gamma \rightarrow l \nu \gamma$ events in which a lepton is lost and $\gamma\gamma$ events in which a photon is lost, a two-step process is used to calculate the expected number of events. First, events in data are selected such that we obtain a very pure sample of one of the aforementioned event types in which there is no lost object. Then, we calculate the fraction of the corresponding events in MC in which an object is not detected, and this fraction is used to scale the photon \Et\ distribution of the data events so as to provide an estimate of this background's photon \Et\ distribution in the signal sample. An exception to this method is the case in which a $\tau$ is lost; due to the difficulty of reliably identifying $\tau$ candidates, this background is estimated purely from MC simulations. Further details on these methods of background prediction can be found in~\cite{LED}, a CDF analysis which used very similar event requirements. We see excellent agreement between the SM predictions and the data in the control regions of $15 < \Etg\ < 40$ GeV (\Zll\ case) and $70 < \Etg\ < 100$ GeV (\Znn\ case).

\begin{table}[!hbtp]
  \begin{center}
    \begin{tabular}{lc}
      \hline
      \hline
     Process        & Events \\
      \hline
      $Z\gamma \rightarrow \nu \bar{\nu}\gamma$ & $52.8 \pm 4.6$ \\
      cosmics & $14.9 \pm 1.4$ \\
      $W \rightarrow e\nu$ & $3.9 \pm 0.8$ \\
      $W \rightarrow \mu/\tau \nu \gamma$ & $1.6 \pm 0.3$ \\
      $W\gamma \rightarrow e\nu\gamma $ & $1.1 \pm 1.1$ \\
      $W\gamma \rightarrow \mu\nu\gamma $ & $1.8 \pm 1.3$ \\
      $W\gamma \rightarrow \tau\nu\gamma $ & $4.5 \pm 1.3$ \\
      $\gamma \gamma$ & $5.3 \pm 1.9$ \\
      \hline
      SM Total & $85.9 \pm 5.6$ \\
      \hline
      Data & 85 \\
      \hline
     \hline
    \end{tabular}
     \caption{SM expected contributions to the $Z\gamma \rightarrow \nu\bar{\nu}\gamma$ candidate sample. Uncertainties shown are systematic only and thus exclude the statistical uncertainties. }
     \label{tab:Znunu_bkgnds}
  \end{center}
    \end{table}

Assuming gauge and Lorentz invariance, eight parameters are needed to describe $Z\gamma$ couplings, denoted by $h^V_{i0}$ where $V$ is either a $Z$ or a $\gamma$ and the index $i$ runs from 1 to 4; these parameters are all zero at tree level~\cite{prd47_1993_4889}. Interaction amplitudes are linear in these parameters. Indices 1 and 2 represent CP-violating terms while indices 3 and 4 represent CP-conserving terms.  We assume CP conservation in these interactions by setting $h^V_{10} =  h^V_{20} = 0$ and we investigate the possibility of non-zero values for $h^V_{30}$ and $h^V_{40}$, corresponding to electric dipole and magnetic quadrupole transition moments~\cite{dipole_quadrupole}. In order to preserve unitarity at large incoming parton center-of-mass energy $\sqrt{\hat{s}}$, an $\hat{s}$-dependent form factor is used to suppress the coupling, constructed as $h^V_i(\hat{s}) = {{h^V_{i0}}\over{(1+\hat{s} /  \Lambda^2)^n}}$, where $n = i$ for $h^V_{30}$ and $h^V_{40}$~\cite{prd47_1993_4889}. The parameter $\Lambda$ describes the predicted energy scale of the new physics that creates anomalous \Zg\ couplings. 

For a given set of anomalous coupling parameter values, we compute a likelihood for the \Etg\ distribution. Hence, we have $\prod_{j=1}^{N} L(x_j | h^V_{i})$, where $x_j$ represents the number of entries in the $j$th of $N$ bins in our \Etg\ distribution and $h^V_{i}$ denotes the coupling parameter being measured (the other three being held fixed at zero). The bin-by-bin likelihood $L$ is simply the Poisson probability of the number of observed entries given the expected number of entries for the value of $h^V_{i}$. This limit method requires a predicted \Et\ distribution for each combination of the four coupling parameters. To create these distributions, we produce \Zg\ MC events at the generator level using the Baur-Berger package~\cite{prd47_1993_4889}. Modeling the particle showering process and detector response in MC separately for every parameter value is computationally impractical. To mimic fully-simulated MC events we first determine the efficiency for a generated event to pass all of the event requirements as a function of generator-level \Etg\ and $|\eta^{\gamma}|$; these functions are derived from a SM MC sample which has used the full simulation of the detector. Due to the correlation between \Etg\ and the $Z$ kinematics, we create and combine separate templates for the cases of central-central, central-forward, and forward-forward lepton pairs, ``central'' denoting $0 < |\eta| < 1.1$ and ``forward'' denoting $1.1 < |\eta| < 2.8$. We then apply this efficiency function to generator-level MC samples to get the expected \Etg\ distributions. The final prediction is the sum of this \Zg\ prediction with the predictions of the non-\Zg\ backgrounds. 

\begin{figure}
 \begin{center}
   \includegraphics[width=1.0 \columnwidth]{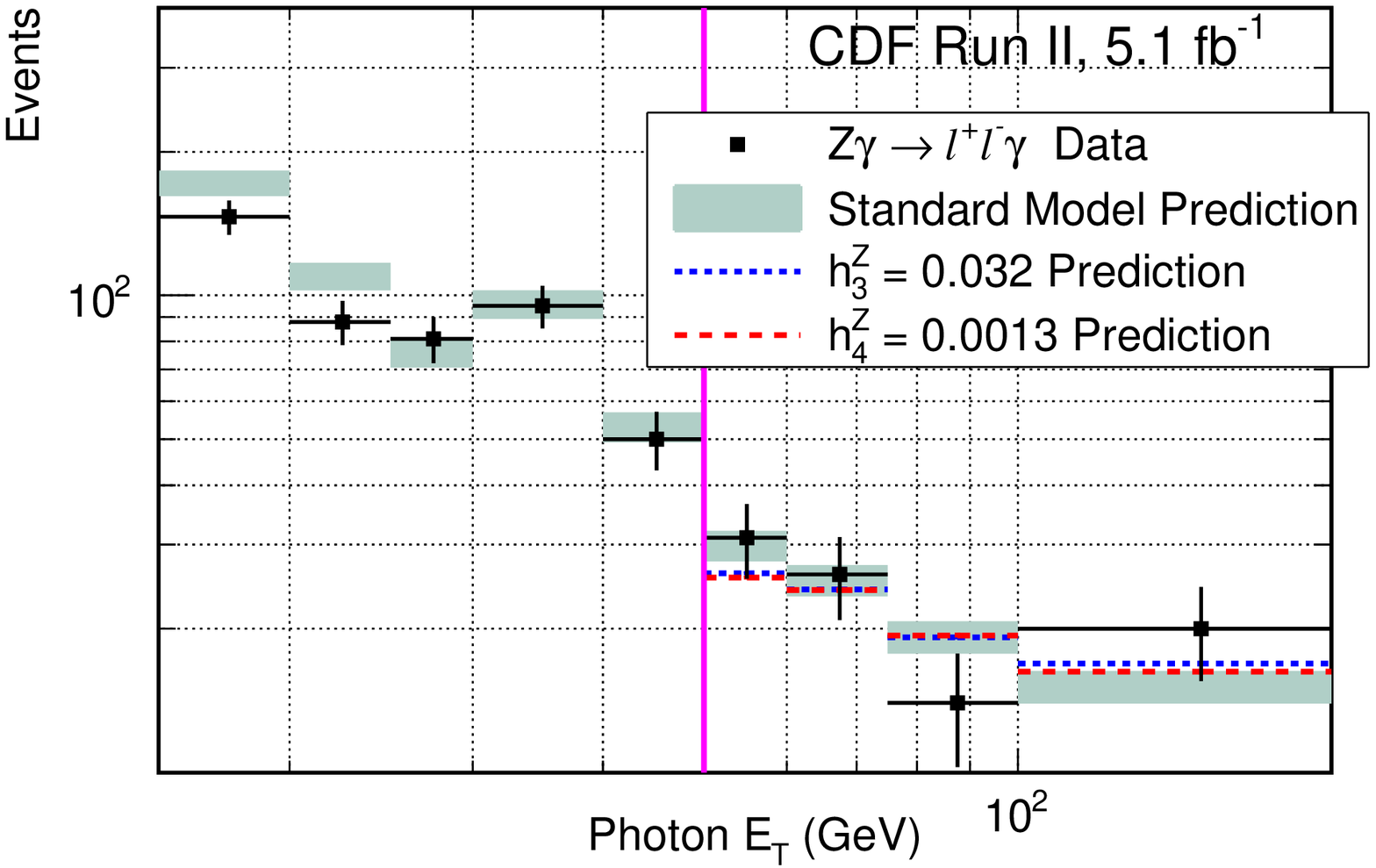}
   \includegraphics[width=1.0 \columnwidth]{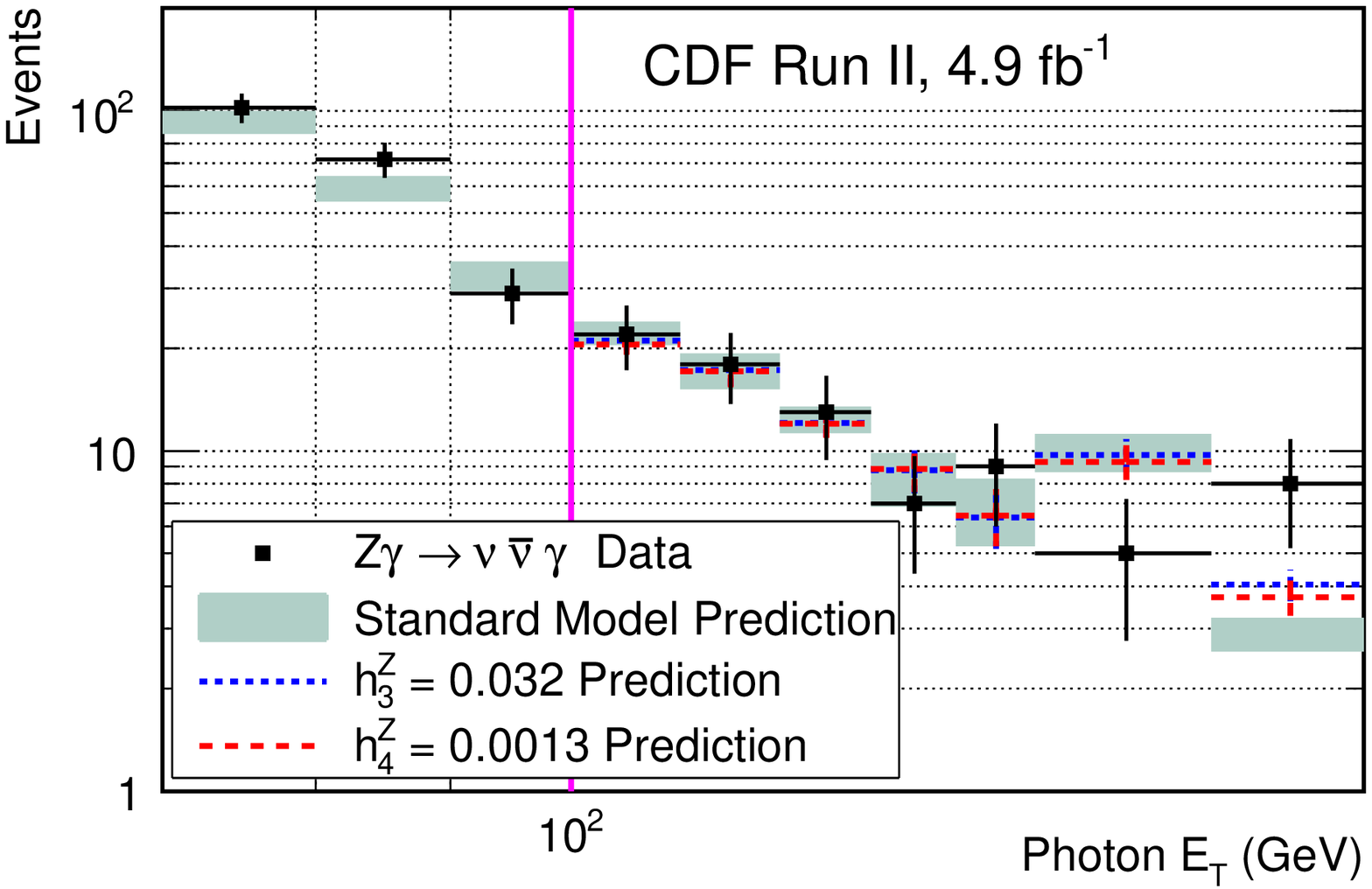}

   \caption{Comparison of the measured \Etg\ distribution with the predicted distributions from both the SM and beyond-the-SM scenarios for $Z \to l^+l^-$ (top) and $Z \to \nu\bar{\nu}$ (bottom) candidate samples, at $\Lambda = 1.5$\ TeV. The beyond-the-SM scenarios chosen here can be excluded at 95\% Bayesian credibility level in each sample. Note the greatest difference in the SM and beyond-the-SM cases is found offscale at \Etg $> 200$ GeV; the lack of data events in this region indicates good agreement with the SM.
   \label{fig:et_comps}}
 \end{center}
 \end{figure}

In Fig.~\ref{fig:et_comps}, for both the \Zll\ and \Znn\ cases, the \Etg\ distributions in data are
compared to the SM prediction and beyond-the-SM predictions; it can be seen that the production of high-\Et\ photons is far more likely in the beyond-the-SM cases compared to the SM case. 
The uncertainty bands shown for the SM predictions illustrate the systematic uncertainties on
those predictions.
These uncertainties are dominated by the 7\% uncertainty on the theoretical \Zg\
cross section~\cite{theoryerror} and the 6\% uncertainty on the luminosity~\cite{lumerror}; the
other sources are the reconstructed photon's energy scale and efficiency, as well as uncertainties
on the number of non-\Zg\ background events. The effect of these systematic uncertainties on the
limits is negligible --- of the order of a couple of percent of the limit values. 

	With the likelihood distribution for a given $h^V_i$, taking a flat Bayesian prior in $h^V_i$ allows us to set Bayesian credibility limits on the parameter. These limits are defined as the values of $h^V_i$ which demarcate the central 95\% of the integral of the likelihood distribution. The resulting allowed ranges for the strength of anomalous couplings are shown in Table~\ref{tab:AGClimits}. The values $\Lambda$ = 1.2 TeV and $\Lambda = 1.5$ TeV have been chosen to allow direct comparisons with earlier CDF~\cite{CDFZllAGC} and D0~\cite{D0ZnunuAGC} results, respectively. We see no evidence for anomalous couplings.  

	In conclusion, we find that the \Etg\ distribution of photons produced in association with $Z$ bosons in both the \Znn\ and \Zll\ decay channels in a data sample corresponding to an integrated luminosity of approximately 5 \invfb\ is consistent with SM couplings. We place 95\% Bayesian credibility limits of $|h_3^{\gamma,Z}| < 0.027 $ and $|h_4^{\gamma,Z}| < 0.0013$ on the CP-conserving \Zg\ couplings at $\Lambda = 1.2$\ TeV and $|h_3^{\gamma,Z}| < 0.022 $ and $|h_4^{\gamma,Z}| < 0.0009$ at $\Lambda = 1.5$ TeV; these are significantly tighter constraints on beyond-the-SM contributions than those provided by previously measured limits.

    \begin{table}
    \begin{center}
    \begin{tabular}{ccc}
     \hline
 \hline
Parameter & ($\Lambda = 1.2$ TeV) & ($\Lambda = 1.5$ TeV)\\
 \hline
           $h_3^Z$         &  $-0.024$, 0.027      & $-0.020$, 0.021                     \\
           $h_4^Z$         &  $ -0.0013$, 0.0013   & $-0.0009$, 0.0009                      \\
 \hline                                                                                            
           $h_3^{\gamma}$  &  $-0.026$, 0.026    & $-0.022$, 0.020              \\
           $h_4^{\gamma}$  &  $ -0.0012$, 0.0013    & $-0.0008$, 0.0008          \\
 \hline
     \hline
    \end{tabular}
       \caption{Allowed ranges (95\% Bayesian credibility limits) of anomalous $Z\gamma$ couplings for $\Lambda=1.2$ and 1.5 TeV using notation from Ref.~\cite{prd47_1993_4889}. Each parameter's limits are set assuming the other three parameters have values fixed at 0. 
     \label{tab:AGClimits}}  
    \end{center}
    \end{table}

We thank the Fermilab staff and the technical staffs of the participating institutions for their vital contributions. This work was supported by the U.S. Department of Energy and National Science Foundation; the Italian Istituto Nazionale di Fisica Nucleare; the Ministry of Education, Culture, Sports, Science and Technology of Japan; the Natural Sciences and Engineering Research Council of Canada; the National Science Council of the Republic of China; the Swiss National Science Foundation; the A.P. Sloan Foundation; the Bundesministerium f\"ur Bildung und Forschung, Germany; the World Class University Program, the National Research Foundation of Korea; the Science and Technology Facilities Council and the Royal Society, UK; the Institut National de Physique Nucleaire et Physique des Particules/CNRS; the Russian Foundation for Basic Research; the Ministerio de Ciencia e Innovaci\'{o}n, and Programa Consolider-Ingenio 2010, Spain; the Slovak R\&D Agency; and the Academy of Finland.


\begin{thebibliography}{99}

\bibitem{CDFZllAGC}
T.~Aaltonen \textit{et al.} (CDF Collaboration), Phys.\ Rev.\ D {\bf 82}, 031103 (2010).

\bibitem{CDFWWAGC}
T.~Aaltonen \textit{et al.} (CDF Collaboration), Phys.\ Rev.\ Lett. {\bf 104}, 201801 (2010). 

\bibitem{D0ZllAGC}
V.M.~Abazov \textit{et al.} (D0 Collaboration), Phys.\ Lett.\ B \textbf{653}, 378 (2007).

\bibitem{D0ZnunuAGC} 
V.M.~Abazov \textit{et al.} (D0 Collaboration), Phys.\ Rev.\ Lett.\ \textbf{102}, 201802 (2009).

\bibitem{D0WZAGC}
V.M.~Abazov \textit{et al.} (D0 Collaboration), Phys.\ Lett.\ B {\bf 695}, 67 (2011).

\bibitem{LEPEWWG}
LEP Electroweak Working Group, LEPEWWG/2006-042.


\bibitem{newZgPhys} G.J.~Gounaris, J.~Layssac, and F.M.~Renard, Phys.\ Rev.\ D \textbf{62}, 073013 (2000).

\bibitem{tgc} For a review see J. Ellison and J. Wudka, Annu. Rev. Nucl. Part. Sci,. \textbf{48}, 1 (1998).


\bibitem{prd41_1990_1476}
U. Baur and E.L. Berger,
Phys. Rev. D \textbf{41}, 1476 (1990).

\bibitem{prd47_1993_4889}
U. Baur and E.L. Berger,
Phys. Rev. D \textbf{47}, 4889 (1993).

\bibitem{gmsb}
  D.~Acosta \textit{et al.}, Phys.\ Rev.\ D \textbf{71}, 031104 (2005).


\bibitem{photonID} F.~Abe \textit{et al.} (CDF Collaboration), Phys.\ Rev.\ D \textbf{52}, 4784 (1995) contains electromagnetic cluster and photon identification variable definitions.

\bibitem{coords} The CDF detector can be described with a coordinate system consisting of $r$, the radial distance from the beam, $\phi$, the angle in azimuth, and $\theta$, the polar angle; the origin is taken to be the geometric center of the detector. The pseudorapidity, $\eta$, is described as a function of $\theta$: $\eta = -$ln$[$tan$(\frac{\theta}{2})]$. Transverse momentum and energy are defined as $\Pt = p \sin \theta$ and $\Et = E \sin \theta$, respectively. 


\bibitem{CDF_run2} The CDF Collaboration, FERMILAB-PUB-96-390-E.

 \bibitem{CDF}
  D.~Acosta \textit{et al.}, Phys.\ Rev.\ D {\bf 71}, 032001 (2005).


\bibitem{SVX} A. Sill \textit{et al}., Nucl. Instrum. Methods A \textbf{447}, 1 (2000).

\bibitem{COT} 
T.~Affolder \textit{et al.}, Nucl.\ Instrum.\ Methods A \textbf{526}, 249 (2004).


\bibitem{showerMax} G. Apollinari,  K. Goulianos, P. Melese, and M. Lindgren,  Nucl.\ Instrum.\ Methods  A \textbf{412}, 515 (1998).

\bibitem{photonIsolation} The additional energy in a cone of $\Delta R < 0.4$ must be less than $ 0.1 \times \Etg$ if $\Etg < 20$ GeV and less than $2+0.02\times(\Etg - 20)$ if $\Etg > 20$ GeV for a photon to pass the isolation selection.

\bibitem{pythia}
T. Sjostrand, S. Mrenna, and P. Skands, J. High Energy Phys. {\bf 0605}, 026 (2006).

\bibitem{MET_defn}
Missing \Et\  (\met ) is defined by $\vec{\met} = -\Sigma_i \Et^i\hat{n}_i$, where $i$ is the calorimeter tower number for 
$|\eta | < 3.6$, and $\hat{n}_i$ is a  unit vector perpendicular to the beam axis and pointing at the $i^{th}$ tower. ($\met = |\vec{\met}|$)


\bibitem{RVM} Chih-Chung Chang and Chih-Jen Lin, \textit{LIBSVM: A Library for Support Vector Machines}, 2001. 

\bibitem{LED} T.~Aaltonen \textit{et al.} (CDF Collaboration), Phys.\ Rev.\ Lett.\ \textbf{101}, 181602 (2008). 

\bibitem{dipole_quadrupole} S.~Abachi \textit{et al.} (D0 Collaboration), Phys.\ Rev.\ D\ {\bf 56}, 6742 (1997).

\bibitem{theoryerror}
U. Baur, T. Han, and J. Ohnemus, Phys. Rev. D \textbf{57}, 2823 (1998). 

\bibitem{lumerror}
D. Acosta et al., Nucl. Instrum. Methods A \textbf{461}, 540 (2001).

\end{thebibliography}
\end{document}